\documentclass[twocolumn, trackchanges]{aastex62}
\usepackage{graphicx}
\usepackage{color,soul}

\received{March 06, 2019}
\accepted{April 19, 2019}
\submitjournal{ApJ}
\shorttitle{Zhou article}
\shortauthors{Zhou et al.}

\begin{document}

\title{High-resolution spectroscopic analysis of a large sample of Li-rich giants found by LAMOST}

\correspondingauthor{Jianrong Shi}
\email{sjr@nao.cas.cn, zhouyutao@nao.cas.cn}

\author{Yutao Zhou}
\affiliation{Key Laboratory of Optical Astronomy, National Astronomical Observatories, Chinese Academy of Sciences, Beijing 100012, PR China}
\affiliation{University of Chinese Academy of Sciences, Beijing 100049, PR China}

\author{Hongliang Yan}
\affiliation{Key Laboratory of Optical Astronomy, National Astronomical Observatories, Chinese Academy of Sciences, Beijing 100012, PR China}
\affiliation{University of Chinese Academy of Sciences, Beijing 100049, PR China}

\author{Jianrong Shi}
\affiliation{Key Laboratory of Optical Astronomy, National Astronomical Observatories, Chinese Academy of Sciences, Beijing 100012, PR China}
\affiliation{University of Chinese Academy of Sciences, Beijing 100049, PR China}

\author{S. Blanco-Cuaresma}
\affiliation{Harvard-Smithsonian Center for Astrophysics, 60 Garden Street, Cambridge, MA 02138, USA}

\author{Qi Gao}
\affiliation{Key Laboratory of Optical Astronomy, National Astronomical Observatories, Chinese Academy of Sciences, Beijing 100012, PR China}
\affiliation{University of Chinese Academy of Sciences, Beijing 100049, PR China}

\author{K. Pan}
\affiliation{Apache Point Observatory and New Mexico State University, P.O. Box 59, Sunspot, NM, 88349-0059, USA.}

\author{Xiaodong Xu}
\affiliation{Key Laboratory of Optical Astronomy, National Astronomical Observatories, Chinese Academy of Sciences, Beijing 100012, PR China}
\affiliation{University of Chinese Academy of Sciences, Beijing 100049, PR China}

\author{Junbo Zhang}
\affiliation{Key Laboratory of Optical Astronomy, National Astronomical Observatories, Chinese Academy of Sciences, Beijing 100012, PR China}

\author{Gang Zhao}
\affiliation{Key Laboratory of Optical Astronomy, National Astronomical Observatories, Chinese Academy of Sciences, Beijing 100012, PR China}
\affiliation{University of Chinese Academy of Sciences, Beijing 100049, PR China}

\begin{abstract}
The discovery of Li-rich giant has cast a new challenge for the standard stellar evolution models, and to resolve this issue, the number of this type object has been rapidly increased because of the development of worldwide surveys these days. Taking advantage of the Large Sky Area Multi-Object Fiber Spectroscopic Telescope survey, 44 newly Li-rich giants are reported, which are confirmed with high-resolution observations. Based on the high-resolution and high signal-to-noise spectra, we derived the atmospheric parameters and elemental abundances with the spectral synthesis method. We performed a detailed analysis of their evolutionary stages, infrared excess, projected rotational velocity ($v\sin i$), and the stellar population. We find that (1) The Li-rich giants concentrate at the evolutionary status of the red giant branch bump, red clump, and asymptotic giant branch; (2) Three of them are fast rotators and none exhibit infrared excess. Our results imply that the origins of Li enrichment are most likely to be associated with the extra mixing in the stellar interior, and the external sources maybe only make a minor contribution. Moreover, various Li-rich episodes take place at different evolutionary stages.

\end{abstract}


\section{Introduction} \label{sec:intro}

Lithium is one of the most important light elements, which is usually considered an indicator of stellar evolution. When a low-mass star evolves to be a cool giant, the standard evolution model predicts that Li will be significantly diluted by a factor of about 60 after the first dredge-up \citep{Iben67a, Iben67b}, Li decreases further because of mixing after the red giant branch (RGB) bump \citep{Sweigart79}. The Li abundances of the majority of RGB stars are less than 1.5 dex, which have been confirmed by the observations \citep{Brown89, Lind09}. However, a few (1-2\%) of giants have been found to exhibit a higher Li abundance than the prediction \citep{Brown89, Charbonnel00}. Some of them even hold a Li abundance that exceeds the value of the interstellar medium ($\sim$3.3 dex; \citealt{Knauth03}). Such stars challenge the standard stellar evolution models.

 Many theories were proposed to account for the Li enhancement, usually classified as the internal and external scenarios. The internal Li production is attributed to the nuclear reaction, $^{3}He(\alpha, \gamma)^{7}Be(e, \mu)^{7}Li$, which is called the Cameron$-$Fowler (CF) mechanism \citep{Cameron71}. The interior of a star is hot enough to produce $^{7}$Be, Li is then created by the $^{7}$Be decay, which requires $^{7}$Be to be quickly transported to the cool region. For the asymptotic giant branch (AGB) stars, the CF mechanism can account for the Li enrichment, because the temperature at the bottom of the convective envelope is high enough to produce the Li-predecessor $^{7}$Be (known as hot bottom burning, HBB). However, the bottom of the envelope in the RGB stars is too cool to produce $^{7}$Be. An extra-mixing is required for conveying material between the outer layer and the inner region where is hot enough to produce $^{7}$Be. A number of mechanisms for the extra-mixing were proposed to trigger Li production, such as thermohaline mixing \citep{Eggleton06}, rotation-induced mixing \citep{Denissenkov04}, and magnetic buoyancy \citep{Busso07}. However, they are still debated.

The external sources are generally regarded as the accretion of the planet, substellar object \citep{Alexander67}, contamination by a more evolved AGB companion that has been through Li-production process \citep{Sackmann99, Kirby16} or accretion of the supernova remnant \citep{Woosley95}. In these cases, some additional traits might accompany the Li-rich phenomenon, such as a high rotational velocity, infrared excess (IR), enrichment in other elements, such as Be \citep{Siess99}, $\alpha$ elements \citep{Woosley95, Jofre15}.

Since the first Li-rich giant was found by \citet{Wallerstein82}, numerous works have been devoted to these peculiar objects. They were discovered at each component of the Galaxy (the disk: \citealt{Balachandran00, Monaco11}, halo: \citealt{Deepak19}, bulge: \citealt{Gonzalez09}, open cluster: \citealt{Anthony-Twarog13, Monaco14}, and globular cluster: \citealt{Ruchti11, Dorazi15, Kirby16}, dwarf galaxies: \citealt{Kirby12}; etc). A consensus amongst the researches is that the Li-rich giants are difficult to find due to the rareness. Majority of Li-rich giants were discovered by the large surveys, such as SDSS \citep{Martell13}, Gaia-ESO \citep{Smiljanic18}, GALAH \citep{Deepak19}, and LAMOST \citep{Li18, Yan18, Zhou18}. In order to investigate the Li-rich giants in detail, high-resolution observation is required. Until now, about 200 Li-rich giants have been identified with high-resolution spectra.

In this paper, we present a high-resolution spectral analysis of the 44 Li-rich giants. All these stars are first reported, which is the largest high-resolution homogeneous sample. These Li-rich candidates were selected from LAMOST survey, and confirmed by follow-up high-resolution observations. It remarkably expands the amount of Li-rich giants. With the spectroscopic information and Gaia astrometry, we investigate the stellar properties and the origin of the Li-rich giant. The paper is organized as follows. In Sect. \ref{sec2}, we gave a brief description of the observation and data reduction. In Sect. \ref{sec3}, we presented the properties of the newly found Li-rich giants. In Sect. \ref{sec4}, we discussed the potential scenarios of Li enhancement based on the evolutionary status, while the summary was proposed in Sect. \ref{sec5}.

\section{Observations and data reduction} \label{sec2} LAMOST is able to take a maximum of 4000 spectra in a single exposure with a limiting magnitude of r=18.5 and a resolution power ($R$) of $\sim$1800 \citep{Cui12, Zhao12}. The wavelength range is 3700 \AA\ $\sim$ 9000 \AA\, which covers the Li resonance line at 6708 \AA. It is of great potential to search for the Li-rich giant in a large amount of spectra. The candidates of the Li-rich giants were picked out from LAMOST spectra, which were based on the equivalent width of Li line at 6708 \AA\ and careful visual inspection to the spectra. Subsequently, these targets were identified by the high-resolution observations. We implemented the follow-up observations with five telescopes, they are 1.8m and 2.4m telescopes at the Lijiang Observatory with $R \sim$ 30,000, 3.5m telescope at Apache Point Observatory (APO) with $R \sim$ 31,500, 2.4m APF telescope located at the Mt. Hamilton station of UCO/Lick Observatory ($R \sim$ 80,000), and 8m Subaru telescope with $R \sim$ 40,000. The mean signal-to-noise ratio (SNR) of the spectra is 80. The details of observation with the basic information (e.g. coordinate, V magnitude, exposure time, etc.) are listed in Table \ref{tab1}.

The spectra were reduced by the revised IDL routines which are primarily designed for the fiber-coupled Cassegrain echelle spectrograph \citep{Pfeiffer98}. The standard reduction procedure includes de-biasing, scattered light subtraction, flat-field correction, order extraction, and wavelength calibration. The continuum fitting, radial velocity measurement and subsequent spectral analysis were performed with iSpec \citep{Blanco-Cuaresma14, Blanco-Cuaresma19}.

\begin{deluxetable*}{cccrccc}
\caption{Observation informations including the LAMOST ID number, J2000 position, V band magnitude, exposure time, telescope and observed time. \label{tab1}}
\tablehead{\colhead{LAMOST ID} & \colhead{RA} & \colhead{DEC} & \colhead{V} & \colhead{$n_{ex} \times t_{exp}$} & \colhead{Telescope} & \colhead{Date-OBS} \\
\colhead{} & \colhead{h:m:s (J2000)} & \colhead{d:m:s (J2000)} & \colhead{mag} & \colhead{} & \colhead{} & \colhead{YYYY/MM/DD}}
\startdata
J005749.84+865043.5 & 00:57:49.84 & +86:50:43.49 & 10.667 & 1800                  & APF-2.4m     & 2016/10/19 \\
J011727.43+461528.3 & 01:17:27.43 & +46:15:28.34 & 9.76   & 2 $\times$ 1800       & Lijiang-2.4m & 2015/10/21 \\
J012741.50+572857.3 & 01:27:41.50 & +57:28:57.27 & 11.851 & 2 $\times$ 1800 + 600 & APO-3.5m     & 2015/09/06 \\
J013000.68+561722.3 & 01:30:00.68 & +56:17:22.33 & 10.519 & 3 $\times$ 1800       & APO-3.5m     & 2014/11/11 \\
J013829.65+840530.5 & 01:38:29.65 & +84:05:30.50 & 10.815 & 1800                  & APF-2.4m     & 2016/10/09 \\
J022740.34+610320.5 & 02:27:40.34 & +61:03:20.53 & 12.107 & 3 $\times$ 1800       & APO-3.5m     & 2015/09/14 \\
J023101.65+625705.3 & 02:31:01.65 & +62:57:05.32 & 12.348 & 4 $\times$ 1800       & APO-3.5m     & 2015/09/30 \\
J023607.44+445007.8 & 02:36:07.44 & +44:50:07.75 & 9.345  & 1200 + 900            & APO-3.5m     & 2015/09/14 \\
J024710.97+432606.0 & 02:47:10.97 & +43:26:06.02 & 11.598 & 3 $\times$ 1800       & APO-3.5m     & 2015/09/14 \\
J032356.98+580337.5 & 03:23:56.98 & +58:03:37.47 & 10.253 & 3 $\times$ 1800   & Lijiang-2.4m & 2016/10/09 \\
J040922.19+641738.3 & 04:09:22.19 & +64:17:38.31 & 12.017 & 3 $\times$ 1800       & APO-3.5m     & 2015/09/30 \\
J054200.63+303722.3 & 05:42:00.63 & +30:37:22.34 & 10.654 & 6 $\times$ 1800       & Lijiang-1.8m & 2015/01/15 \\
J055908.81+120339.7 & 05:59:08.81 & +12:03:39.68 & 11.149 & 2 $\times$ 1800       & APO-3.5m     & 2013/10/21 \\
J060649.27+212504.9 & 06:06:49.27 & +21:25:04.90 & 11.767 & 3 $\times$ 1800       & APO-3.5m     & 2015/12/08 \\
J061210.83+230918.9 & 06:12:10.83 & +23:09:18.93 & 11.135 & 2 $\times$ 1800 + 800 & APO-3.5m     & 2015/09/30 \\
J062853.52+042445.1 & 06:28:53.52 & +04:24:45.08 & 9.724  & 4 $\times$ 1800       & Lijiang-1.8m & 2017/01/15 \\
J063640.99+030112.3 & 06:36:40.99 & +03:01:12.33 & 10.171 & 3 $\times$ 1800       & Lijiang-2.4m & 2017/01/17 \\
J064934.47+170424.2 & 06:49:34.47 & +17:04:24.15 & 11.178 & 2 $\times$ 1800       & APO-3.5m     & 2013/10/21 \\
J071813.82+500452.6 & 07:18:13.82 & +50:04:52.61 & 9.719  & 5 $\times$ 1800       & Lijiang-1.8m & 2017/02/15 \\
J072619.82+295808.2 & 07:26:19.82 & +29:58:08.15 & 9.524  & 4 $\times$ 1800       & Lijiang-1.8m & 2017/02/14 \\
J072840.88+070147.4 & 07:28:40.88 & +07:01:47.36 & 9.806  & 3 $\times$ 1800       & Lijiang-1.8m & 2017/03/14 \\
J073954.88+213949.7 & 07:39:54.88 & +21:39:49.65 & 9.826  & 4 $\times$ 1800       & Lijiang-1.8m & 2017/01/14 \\
J074051.22+241938.3 & 07:40:51.22 & +24:19:38.27 & 10.885 & 1 $\times$ 1800       & APO-3.5m     & 2015/12/08 \\
J074738.59-004430.7 & 07:47:38.59 & -00:44:30.69 & 9.85   & 3 $\times$ 1800       & Lijiang-1.8m & 2017/03/16 \\
J085802.95+122935.2 & 08:58:02.95 & +12:29:35.15 & 9.893  & 3 $\times$ 1800   & Lijiang-1.8m & 2014/11/11 \\
J085929.54+005654.2 & 08:59:29.54 & +00:56:54.15 & 10.367 & 4 $\times$ 1800       & Lijiang-1.8m & 2015/01/21 \\
J102823.20-044851.0 & 10:28:23.20 & -04:48:50.98 & 9.798  & 3 $\times$ 1800       & Lijiang-1.8m & 2017/03/10 \\
J103249.02+143714.8 & 10:32:49.02 & +14:37:14.83 & 9.229  & 4 $\times$ 1800       & Lijiang-1.8m & 2017/02/13 \\
J110236.56+133610.3 & 11:02:36.56 & +13:36:10.27 & 10.175 & 4 $\times$ 1800       & Lijiang-1.8m & 2017/01/14 \\
J122234.29+321817.2 & 12:22:34.29 & +32:18:17.16 & 9.622  & 3 $\times$ 1800       & Lijiang-1.8m & 2017/03/10 \\
J122525.23+071638.0 & 12:25:25.23 & +07:16:38.00 & 10.336 & 3 $\times$ 1800       & Lijiang-1.8m & 2017/01/16 \\
J131628.98+431553.0 & 13:16:28.98 & +43:15:53.04 & 9.429  & 4 $\times$ 1800       & Lijiang-1.8m & 2017/02/10 \\
J132315.71+034347.4 & 13:23:15.71 & +03:43:47.36 & 9.787  & 4 $\times$ 1800       & Lijiang-1.8m & 2017/02/15 \\
J143038.38+532629.5 & 14:30:38.38 & +53:26:29.49 & 9.459  & 2 $\times$ 1800       & Lijiang-1.8m & 2017/03/16 \\
J144737.25+142541.7 & 14:47:37.25 & +14:25:41.68 & 9.355  & 2 $\times$ 1800       & Lijiang-1.8m & 2017/03/10 \\
J153707.04+182421.0 & 15:37:07.04 & +18:24:21.04 & 9.788  & 4 $\times$ 1800       & Lijiang-1.8m & 2017/03/16 \\
J154330.58+003038.9 & 15:43:30.58 & +00:30:38.93 & 10.809 & 3 $\times$ 1200       & APO-3.5m     & 2015/05/20 \\
J161035.91+331604.8 & 16:10:35.91 & +33:16:04.84 & 11.558 & 300                   & Subaru-8m    & 2017/02/17 \\
J162103.81+363017.5 & 16:21:03.81 & +36:30:17.54 & 9.855  & 1800 + 900            & APO-3.5m     & 2015/04/06 \\
J170124.77+144913.0 & 17:01:24.77 & +14:49:13.00 & 9.973  & 2 $\times$ 1200       & APO-3.5m     & 2015/04/06 \\
J214122.20+482643.5 & 21:41:22.20 & +48:26:43.47 & 10.758 & 1800                  & APF-2.4m     & 2016/10/11 \\
J225902.66+054256.2 & 22:59:02.66 & +05:42:56.15 & 9.971  & 2 $\times$ 1800       & Lijiang-2.4m & 2015/10/22 \\
J232834.22+573418.5 & 23:28:34.22 & +57:34:18.45 & 10.596 & 2 $\times$ 1800       & Lijiang-2.4m & 2015/10/19 \\
J234211.40+641429.1 & 23:42:11.40 & +64:14:29.07 & 10.346 & 1800                  & APF-2.4m     & 2016/10/09 \\
J235043.31+361105.7 & 23:50:43.31 & +36:11:05.65 & 10.205 & 3 $\times$ 1800       & Lijiang-2.4m & 2015/10/21
\enddata
\end{deluxetable*}

\section{Spectral analysis} \label{sec3}
\subsection{Atmospheric parameters and abundances}
We derive the atmospheric parameters (i.e. effective temperature $T_\mathrm{eff}$, surface gravity log $g$, metallicity and microturbulent velocity) by fitting the observed spectra to the synthetic spectra computed using iSpec with the radiative transfer code SPECTRUM \citep{Gray94}. The MARCS 1D model atmospheres \citep{Gustafsson08} and the Gaia-ESO Survey version 5 atomic line list \citep{Heiter15} are adopted. A least-square algorithm minimization is executed to reduce the differences between the observed and synthetic spectra for a delimited set of regions, the parameters are obtained when the process reaches convergence after several iterations. The selection of lines was done by line-by-line differential analysis in the NARVAL solar spectrum, and the solar abundances have been adopted from \citet{Grevesse07}. The good lines are expected to better reproduce the solar spectrum when we fix the atmospheric parameters to the solar reference, and we selected these lines with a difference of abundance within $\pm$ 0.05 dex given the solar spectrum.

After determining the atmospheric parameters, Li abundances were obtained by keeping the stellar parameters fixed and using the Li resonance line at 6708 \AA. The line list includes the hyperfine structure and isotopic splitting of the Li resonance line. The local thermodynamic equilibrium (LTE) abundances of Li were derived using the SPECTRUM. It is noted that strong Li lines suffer a large non-LTE effect, thus, we also derived the non-LTE results based on the Li atomic model of \citet{Shi07}.

The covariance errors were returned as the uncertainties of atmospheric parameters by the nonlinear least-squares fitting algorithm in iSpec. In the case of the Li abundances, the errors of abundances are the quadratic sum of uncertainties of the atmospheric parameters. The atmospheric parameters and Li abundances are presented in Table \ref{tab2}.

\begin{figure}
 \centering
   \includegraphics[width=8.5cm]{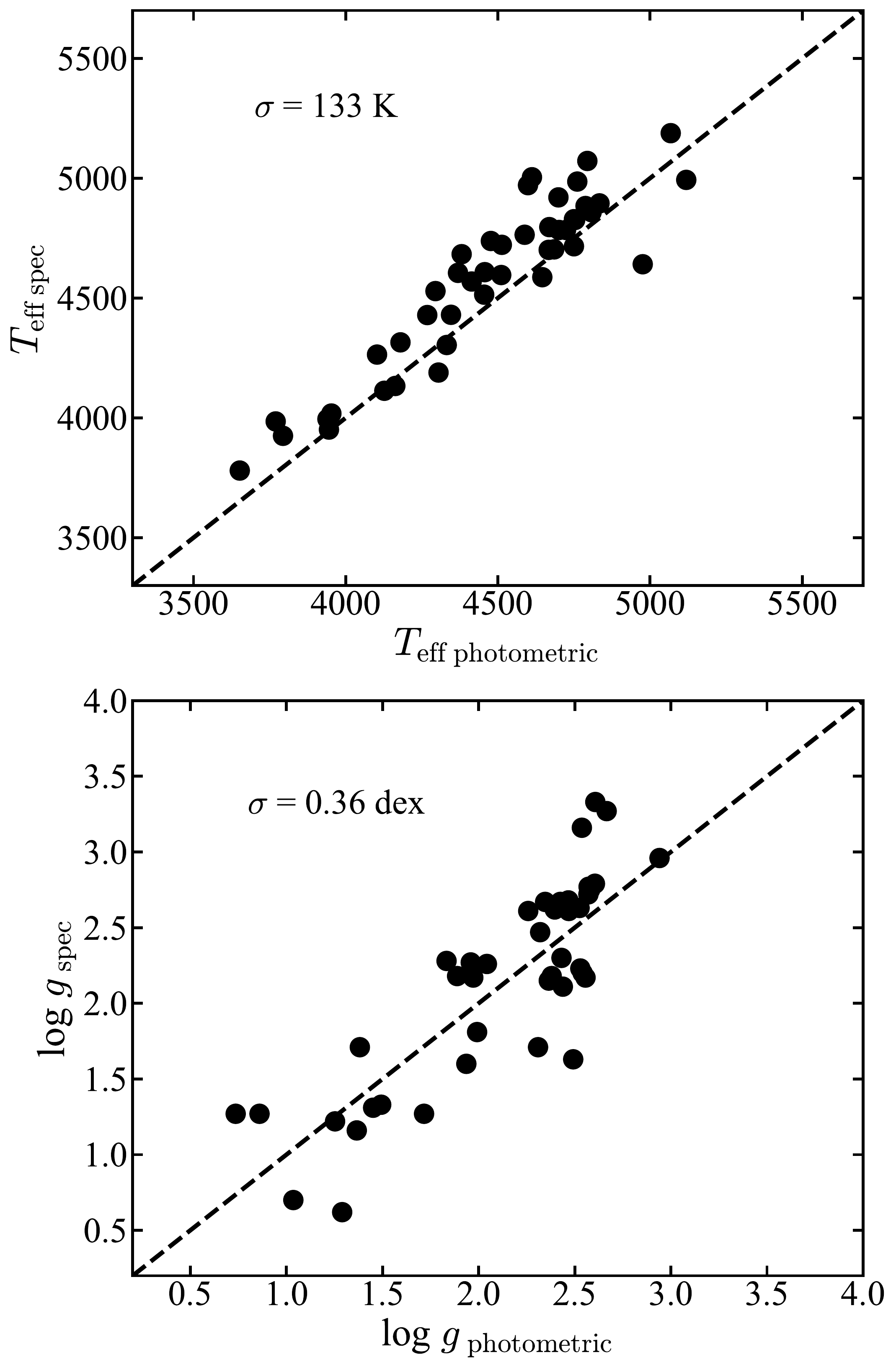}
      \caption{The comparison between methods of the spectroscopy and photometry, the dash is the diagonal line.}
   \label{fig1}
\end{figure}

For an independent check of the spectroscopic parameters, we calculated the photometric $T_\mathrm{eff}$ using \citet{Alonso99} color-temperature empirical relations of intrinsic color (V-K). The extinction was provided by the 3D dust map of \citet{Green15}, which is based on the Pan-STARRS and 2MASS photometric data. It is noted that the extinction of one star is not available by \citet{Green15}, which was adopted from \citet{Green18}. The second data release of Gaia has provided high precision astrometric information including proper motion and parallax for an unprecedented number ($\sim$1.3 billion) of stars (DR2; \citealt{Gaia16, Gaia18}). For all of our targets, the uncertainties of parallaxes are less than $\sim$10$\%$ from Gaia DR2, the distances were derived from \citet{Bailer-Jones18}, which is a probabilistic approach by adopting a prior of the expected distribution of all stars in Gaia measurements. The photometric surface gravity was then derived through the equation:
\begin{eqnarray}
\mathrm{log}(\frac{g}{g_{\odot}}) = \mathrm{log}(\frac{M}{M_{\odot}}) + 4\mathrm{log}(\frac{T_\mathrm{eff}}{{T_\mathrm{eff}}_{\odot}}) + 0.4(M_{bol} - {M_{bol}}_{\odot}) \nonumber
\end{eqnarray}
where values of $M_{\odot}$, $T_\mathrm{eff\odot}$, $M_{bol\odot}$ were assumed to be 4.74, 5772, 4.44 for the Sun, respectively. To derive the absolute bolometric magnitude ($M_{bol}$), the bolometric correction was calculated using the empirical relation of \citet{Alonso99} for the giants. The luminosity was subsequently obtained through the $M_{bol}$. The comparisons between spectroscopic and photometric approaches are shown in Figure \ref{fig1}. Although $T_\mathrm{eff}$ show a minor offset in the comparison diagram, the log $g$ demonstrate one to one around the diagonal lines. Stellar masses and ages were estimated by the PARAM 1.3 \footnote{\url{http://stev.oapd.inaf.it/cgi-bin/param_1.3}}, which is a Bayesian PARSEC-isochrones fitting code. It is a statistical method that matches the observed data to the model parameters from the stellar evolution. The stellar masses and ages are listed in Table \ref{tab2}.

\subsection{Rotation and infrared excess}
\begin{figure}
 \centering
  \includegraphics[width=8.5cm]{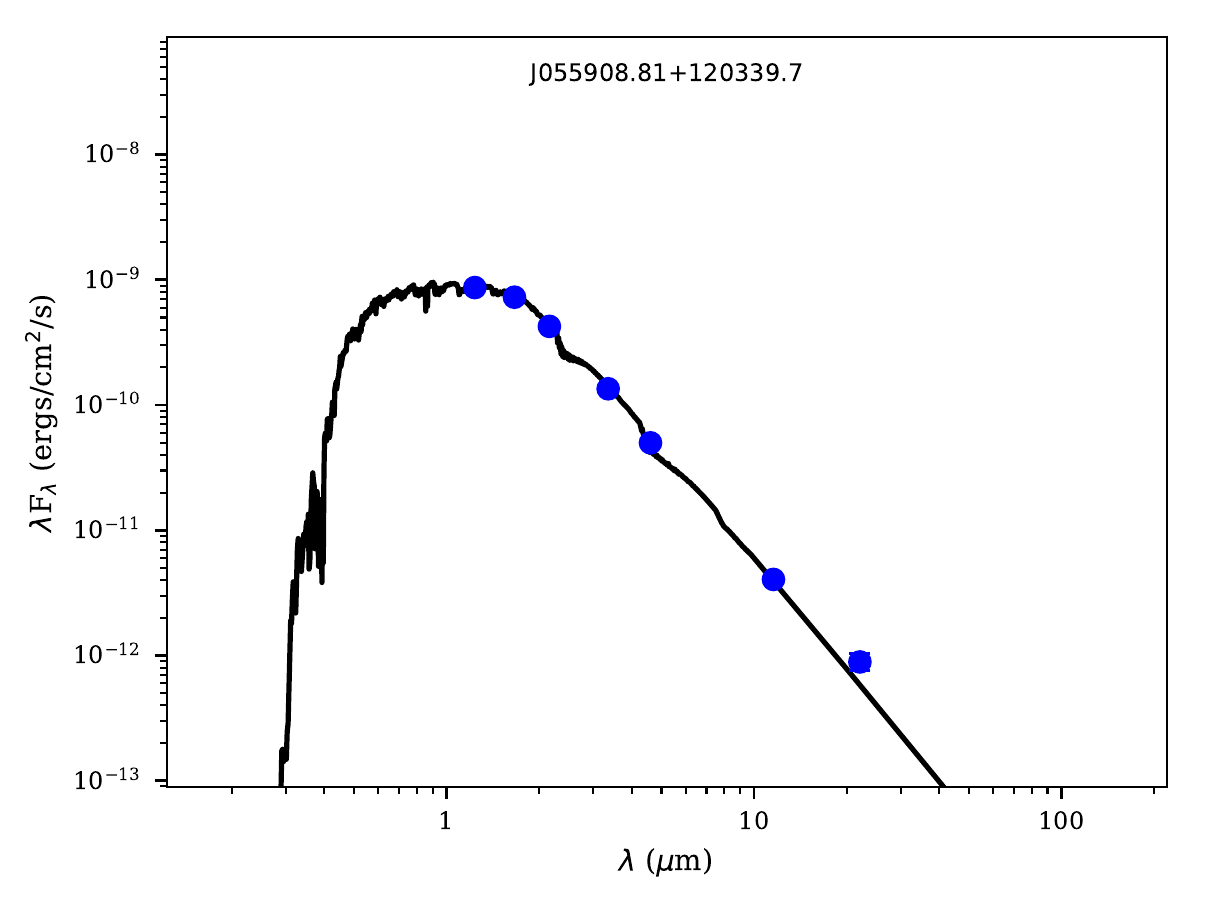}
   \caption{SED for star J055908.81+120339.7 with the largest $\chi_{[3.4],[22]}$.}
  \label{fig2}
\end{figure}

Fast rotation is an expected phenomenon of the external origin that enhances Li abundance, which is supported by some studies. For instance, \citet{Drake02} found that a high ratio (almost 50\%) of fast rotators are Li-rich giants, the projected rotational velocity ($v\sin i$) of these objects are greater than 8 $\mathrm{km\,s}^{-1}$. The fast rotation might be a result of the planet engulfment \citep{Carlberg09} or the tidal interaction with a binary companion \citep{Denissenkov04}. The high projected rotational velocity is generally considered a key element in investigating the origin of Li enhancement. The external line broadening is due to instrumental broadening, macroturbulence velocity, and projected rotational velocity \citep{Bruntt10}. The instrumental broadening can be measured from the Th-Ar emission lines, and the macroturbulence velocity can be derived from an empirical relation based on $T_\mathrm{eff}$ and log $g$ \citep{Hekker07}. Except for the projected rotational velocity, all parameters for line broadening were fixed, $v\sin i$ was obtained by fitting the profile of the selected iron lines. For our targets, we derived the projected rotational velocity by fitting the neutral iron lines at 6703, 6705, 6726, 6733, and 6750 \AA, and the adopted mean values are shown in Table \ref{tab2}.

Infrared excess is also proposed to link the high Li abundance due to the mass-loss event or circumstellar shell ejection \citep{dela96, dela97}, and a large infrared excess can be easily recognized in the spectral energy distributions (SEDs). Based on the WISE and 2MASS photometry, we investigated the SED through the comparison between the observed fluxes and those from Kurucz models in the near-IR regions \citep{Castelli04}. It is found that the observed fluxes from H band to W4 band are coincident with the theoretical predictions, and there are no any significant excesses presented in the SEDs, an example of SED fitting is shown in Figure \ref{fig2}. In addition, if the giant suffers an intensive mass-loss event, an asymmetric profile of the H$\alpha$ absorption line would appear \citep{Meszaros09}, and no such feature have been detected in any star of our sample.

A more subtle method was proposed to identify small IR excess by \citet{Trilling08} and \citet{Mizusawa12}, which was used by \citet{Rebull15}. Following that work, we calculated the $\chi_\mathrm{[3.4],[22]}$ using the WISE data, which is an ideal metric used to identify infrared excess (Table \ref{tab2}). The photometric measurement in each band is labeled as "A" corresponding to high quality data (SNR $\ge$ 10) for most of our targets. The definition of $\chi_\mathrm{[3.4],[22]}$ is as follows:
\begin{eqnarray}
\chi_\mathrm{[3.4],[22]}=\frac{([3.4]-[22])_\mathrm{observed}-([3.4]-[22])_\mathrm{predicted}}{\sigma_\mathrm{([3.4]-[22])}} \nonumber
\end{eqnarray}
where [3.4] and [22] indicate the magnitudes at bands of 3.4 $\mu$m and 22 $\mu$m from WISE \citep{Cutri13}, respectively. Here, $\chi > 3$ is considered as an indicator of small infrared excess. The predicted difference between W1 and W4 is normally expected to be 0 for K giants (for more details see \citealt{Rebull15}). None of subtle IR excess was detected by means of $\chi_\mathrm{[3.4],[22]}$, which is consistent with the result of the SEDs.

\begin{figure}
 \centering
   \includegraphics[width=8.5cm]{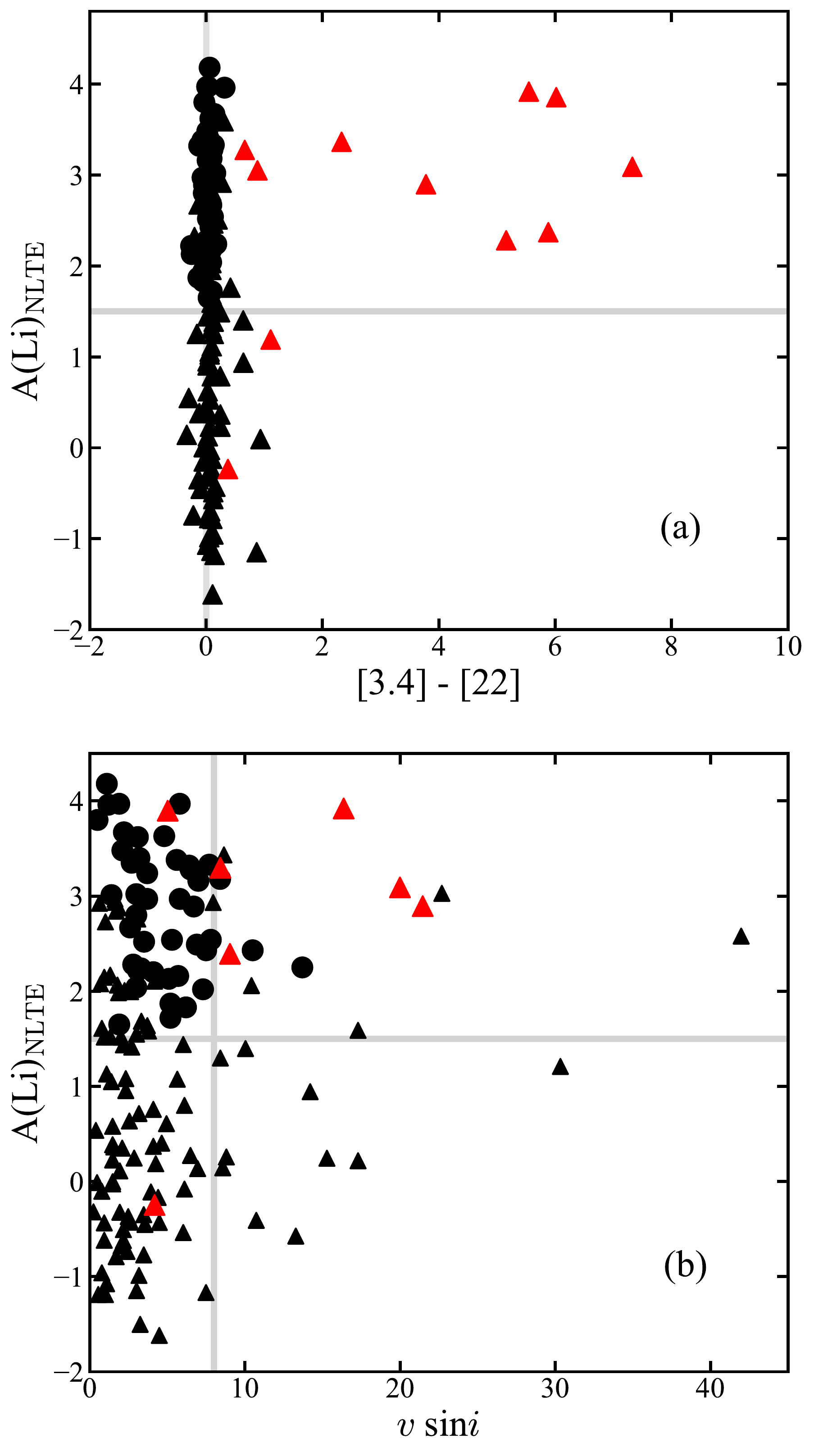}
      \caption{(a) A(Li)$_\mathrm{NLTE}$ function of [3.4] - [22]. The vertical line indicates that the value of [3.4] - [22] = 0 is expected for the photosphere of K giants. (b) Li abundance versus $v\sin i$. The gray line at $v\sin i$ = 8 $\mathrm{km\,s}^{-1}$ is the division between fast and slow rotators. The horizontal line is the conventional standard to define the Li-rich at A(Li)=1.5. Our sample stars are denoted as the circles, and the triangles indicate the sources collected from \citet{Rebull15}. The red targets mean the stars exhibit infrared excess.}
 \label{fig3}
\end{figure}

In Figure \ref{fig3}, we presented an analogous plot of figure 18 and 19 of \citet{Rebull15} for our program stars to examine the relations of Li abundance to $v\sin i$ and infrared excess. \citet{Rebull15} discovered that about ten of their sample stars are fast rotators, and half of them exhibit IR excess. In our sample, only three Li-rich giants are identified as fast rotators, and the ratio of the fast rotator (6.8$\%$) is slightly lower than that (12$\%$) of \citet{Rebull15}. Although \citet{Rebull15} found that a half of their Li-rich giants with fast rotation exhibit infrared excess, none of our fast rotators show the infrared excess.

\subsection{Stellar population}
Different stellar populations have distinct origins, thus, Li abundance might evolve through different paths for the Galactic thick- and thin-disk objects. \citet{Fu18} found that more Li is produced in the Galactic thin-disk phase than in the thick-disk phase, which implies different histories of Li enrichment in these two populations. To investigate the Li-rich giants in different populations, we separated our Li-rich giants into different stellar populations based on their kinematics. We calculated the spatial velocity components (UVW) with respect to the local standard of rest using Astropy package \footnote{https://www.astropy.org}\citep{Astropy13, Astropy18}. The standard solar motion (U, V, W) = (7.01, 10.13, 4.95) km s$^{-1}$ from \citet{Huang15} was adopted. Following \citet{Reddy06}, the spatial velocities with the probabilities of each population are presented in Table \ref{tab2}. Except for the star J161035.91+331604.8, all the stars show that the probability of the thin disk is greater than 70$\%$ \citep{Adibekyan12}, which indicate that most of our sample stars belong to the thin disk. As suggested by \citet{Fu18}, the Galactic thin disk both has higher Li abundance and higher Li enrichment compared to the thick disk. This might lead to a high frequency of Li-rich giant in the thin disk under the current definition of Li-rich.

\subsection{Evolutionary stage}
\begin{figure*}
 \centering
 \includegraphics[width=17cm]{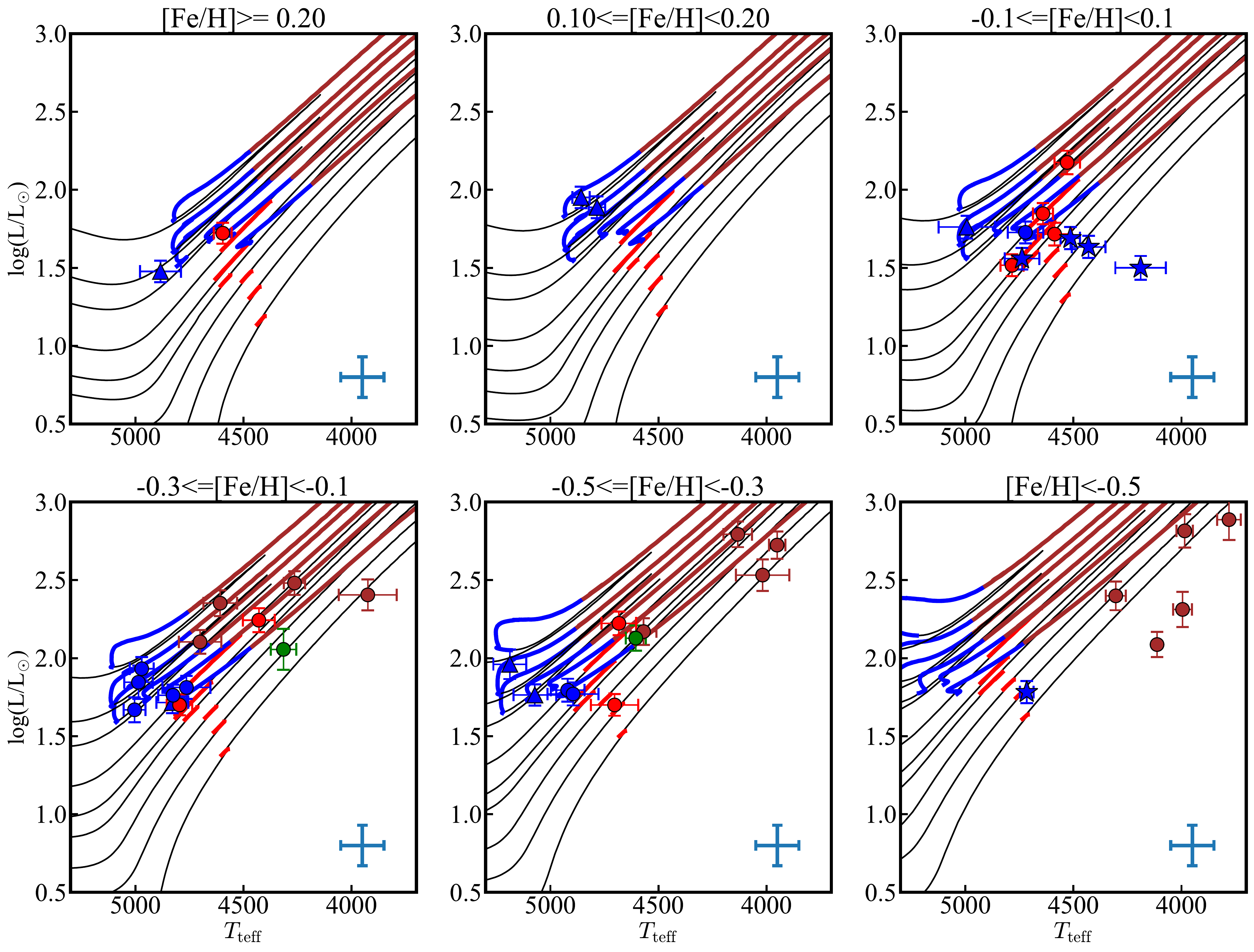}
 \caption{The positions of 44 Li-rich giants displayed in the HR diagrams. The PARSEC evolutionary tracks \citep{Bressan12} are adopted with masses of 0.8, 1.2, 1.4, 1.5, 1.7, 2.0, 2.2 and 2.6 $M_\mathrm{\odot}$. According to the distribution of the metallicity, the stars are divided into six metallicity bins and the tracks are plotted with [Fe/H] = 0.34, 0.14, -0.02, -0.18, -0.40 and -0.58. The portions of PARSEC evolutionary tracks in colors of red, blue and brown denote the phase of RGB bump, RC and AGB, respectively. While, the circles in red, green and brown represent the RGB bump stars, stars evolved past the RGB bump and AGB stars. The RC stars are denoted in blue with various kinds of symbols. The stars and triangles are the RCs determined by only one of the approaches: data-driven method of \citet{Ting18} or HR diagram, and the circles indicate that the RCs were consistently classified with the two methods. The typical errors of luminosity and $T_\mathrm{eff}$ are shown with 100 K and 0.13 dex at the bottom right of each panel.}
 \label{fig4}
\end{figure*}

 Lithium enhancement is proposed to link some evolutionary stages \citep{Charbonnel00, Kumar11, Smiljanic18}. It is important to investigate the evolutionary stage to reveal the origin of Li excess. To clearly show each Li-rich giant in the Hertzsprung-Russel (HR) diagram (Figure \ref{fig4}), we displayed the Li-rich giants on the PARSEC evolutionary tracks with masses ranging from 0.8 to 2.6 $M_{\odot}$ at six panels divided by the metallicity. As shown in Figure \ref{fig4}, the Li-rich giants locate across from RGB to AGB. Most of them are grouped in the region of the RGB bump, RC and AGB, and only a few reside above the RGB bump.

The stars are difficult to confidently distinct from RC to RGB bump in the HR diagram by the spectroscopic parameters alone, while the asteroseismic parameters including average period spacing ($\Delta P$) and the frequency separation ($\Delta \nu$) can help. \citet{Bedding11} suggested that the stars with $\Delta P \ge$ 150 and $\Delta \nu \le$ 5 is a valid criterion to separate RC giants with He-core burning from RGB giants with H-shell burning. Unfortunately, our targets did not have the available asteroseismic information from the time-series observations, such as $Kepler$, TESS. Recently, \citet{Ting18} developed a data-driven method to obtain the asteroseismic data, and found that RC stars could be identified with a low RGB contamination rate ($\sim 9 \%$) for the LAMOST DR3 targets. We crossed match with the above catalog, 15 of our Li-rich giants have been analyzed by the data-driven method, and \citet{Ting18} classified 13 of them to be RC stars.

 As mentioned above, we determined the evolutionary stage by the positions of stars in HR diagrams. In Figure \ref{fig4}, we assign the different symbols to indicate the different evolutionary stages for the Li-rich giants. Although most of them are consistently classified by both the data-driven method and the position on the HR diagram, some exceptions are encountered. Among the RC stars, five stars clearly locate at the RGB region, while they are identified as RC stars according to \citet{Ting18}. Considering the errors of stellar parameters (i.e. $T_\mathrm{eff}$ and luminosity), these targets are too far away from the RC areas (at the top right and the last panels). On one hand, the discrepancy might be caused by a large uncertainty of reddening, because these Li-rich giants are close to the Galactic plane ($\sim$300 pc). On the other hand, they are possibly mis-classified as RC stars due to the RGB contamination of \citet{Ting18}.

\section{Discussions} \label{sec4}
\begin{figure*}
 \centering
   \includegraphics[width=13cm]{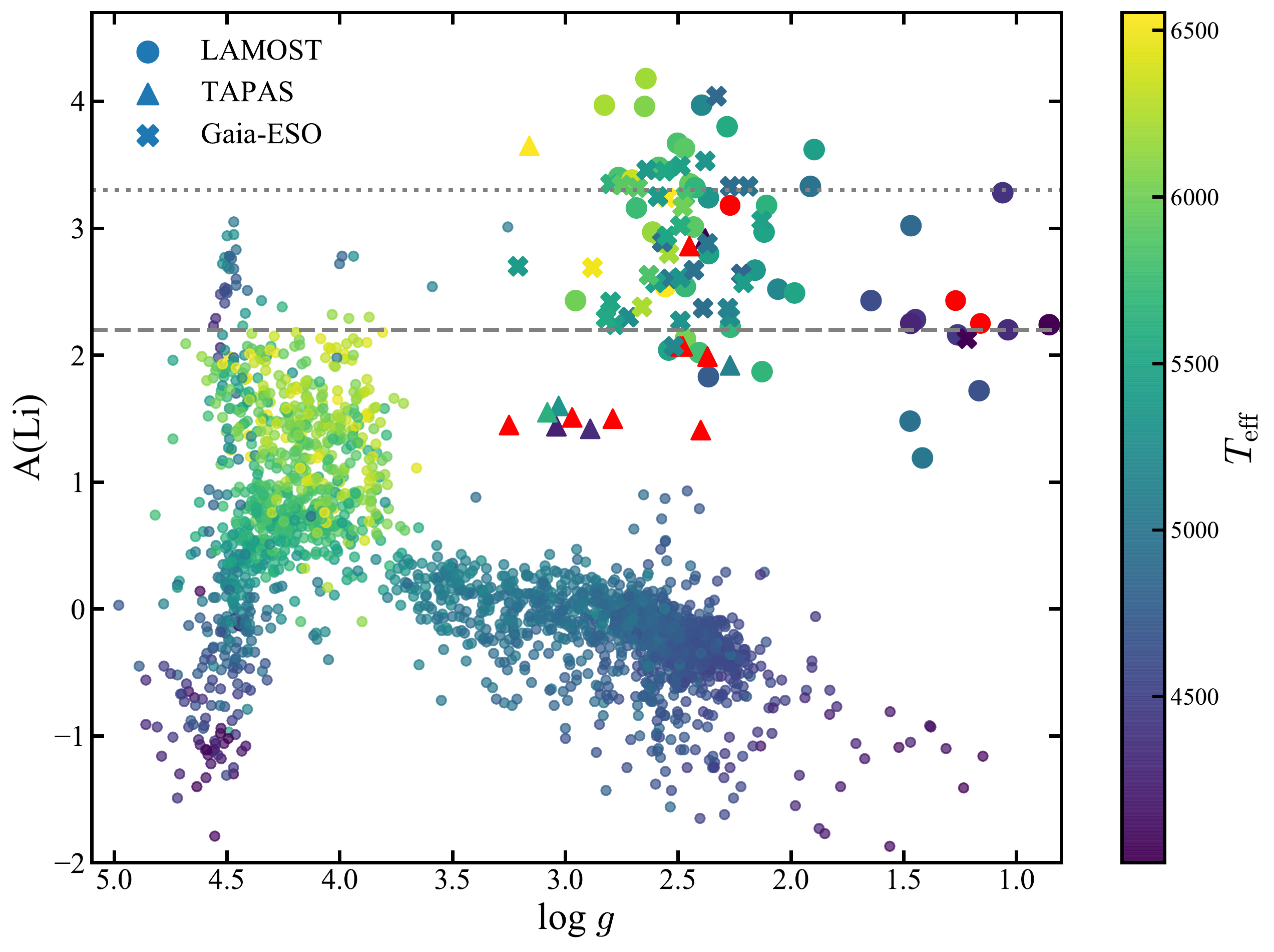}
      \caption{The Li abundances versus log $g$ for our targets (circles) with the Li-rich giants discovered by the Gaia-ESO (crosses) and TAPAS (triangles). The small points are the stars with Li abundance from the Gaia-ESO DR3. The red circles are the fast rotators in our sample, while the red triangles indicate the stars with companions from TAPAS. The dotted line implies the Li abundance at the level of 2.2 dex predicted by the external models of \citet{Aguilera16}, while the dashed line indicates the ISM value at A(Li) = 3.3 dex. The effective temperature is scaled to the colors.}
       \label{fig5}
\end{figure*}

\subsection{external and internal Li-rich mechanism}

Two main mechanisms that included internal and external origins are attributed to the Li enhancement. The key factors to distinct these two hypotheses mainly depend on the associated phenomena with the Li overabundance (e.g. infrared excess, fast rotation, other chemical peculiarities). When a star climbs to RGB, the radius of the star increases, and it is most likely to engulf or accrete the nearby companions \citep{Siess99}. One of the expected effects of companion accretion is the ejection of the shell, and there is subsequent emission in the infrared. Using the Infrared Astronomy Satellite (IRAS) colors, \citet{dela96} suggested that the mass-loss with the detection of IR excess would associate with the Li-rich phase. However, \citet{Rebull15} analyzed the connection between the Li overabundance and IR excess in a sample of $\sim$ 80 Li-rich giants, and did not find any obvious correlation (also see \citealt{Kumar15}). The Li abundance against the IR excess is presented in Figure \ref{fig3}, and we found that our Li-rich giants exhibit neither a large IR excess in SEDs nor a small IR excess in [22] $\mu$m.  Similarly, \citet{Kumar15} found that none of the Li-rich giants have IR excess based on a sample of 2000 K giants, and concluded that IR excess is not correlated with the Li-rich phenomenon.

In addition, the external source to enrich Li would probably increase the projected rotational velocity of stars due to the orbital angular momentum transferred from the external source. \citet{Drake02} claimed that the rate of Li-rich RGB stars would raise to 50\% if $v\sin i \ge$ 8 $\mathrm{km\,s}^{-1}$ is considered. \citet{Carlberg12} made a systematic investigation on the relationship of Li overabundance and the rotational velocity, and they divided their sample into two groups by the standard of $v\sin i \ge$ 8 $\mathrm{km\,s}^{-1}$, and found that the average of Li abundance of the fast group is $\sim$ 1.0 dex higher than that of the slow one. This may indicate that high rotational velocity plays a key role in Li enhancement. In the case of our sample, the projected rotational velocities are mainly around 3 $\sim$ 5 $\mathrm{km\,s}^{-1}$, which are the typical values for the K giants \citep{Gray89}. Only three of them are detected with a projected rotational velocity greater than 8 $\mathrm{km\,s}^{-1}$.

 \citet{Rebull15} suggested that a shell ejection (i.e. a large IR excess shown in the SED) might associate not only with Li enhancement but also with fast rotation. If it is the case, the IR excess would be observed for these fast rotators. However, we can not detect any IR excess for the three fast rotators.

The phenomena of IR excess and fast rotation are only the speculation inferred by external episodes. It is important to test whether Li-rich giants have companions. By monitoring the variation of radial velocity, a sample of Li-rich giants was confirmed with companions by using the Tracking Advanced Planetary Systems (TAPAS) \citep{Adamow18}. For Li-rich giants with Li abundance lower than 2.2 dex (Figure \ref{fig5}), they can be explained by the engulfment episode of \citet{Aguilera16}. In the model of \citet{Aguilera16}, the maximum Li abundance of RGB star can reach $\sim$ 2.2 dex after the digestion of a substellar companion with 15 Jupiter mass if the mixing induced by the companion is not taken into account. Although two of the three fast rotators of our sample have Li abundances around the predicted value of \citet{Aguilera16}, a number of our stars have Li abundances higher than 2.2 dex, 14 of them even show the Li abundances exceeding the Interstellar Medium (ISM) value ($\sim$3.3 dex). The external mechanism is unlikely to interpret all of our program stars. Considering the low rates of fast rotator and IR excess, the external mechanism might make a minor contribution to the Li-rich giants.

Other than the direct Li contamination by a companion, \citet{Denissenkov04} proposed that the interaction with a companion could trigger the extra mixing. This process is not limited to a specific evolutionary stage, which can account for the Li-rich giants found anywhere along the RGB. In contrast, the mechanism of internal mixing is usually bonded with some particular evolutionary stages. However, the trigger of extra mixing is still not clear (see discussion in section 4.2).

 Based on the Li-rich giants from Gaia-ESO iDR4, \citet{Casey16} reported that most of their stars locate below the RGB bump. They thought that these objects are difficult to be explained through internal mixing, as this process is believed to operate at or past the RGB bump. However, \citet{Smiljanic18} re-investigated the Li-rich giants from Gaia-ESO survey including the sample of \citet{Casey16} benefited from the high-quality parallaxes from Gaia and the revised stellar parameters of iDR5, and they found that the Li-rich giants of Gaia-ESO locate near the RGB bump, He-core burning stages, and early-AGB. A similar result was found based on our more extensive sample, which suggests that most of our Li-rich giants are likely connected to the internal mechanisms.

\subsection{Li-rich phase in HR diagram}
 Different explanations for the Li enrichment may be needed for different stellar evolutionary stages. Several scenarios have been proposed for the Li-rich giants at the evolutionary stage of the RGB bump. The prevailing cause of extra mixing is assumed to be thermohaline instabilities \citep{Eggleton06} or magnetic buoyancy \citep{Busso07}. \citet{Yan18} also speculated the combination of these two physical causes for the most Li-rich giant with A(Li) $\sim$ 4.5, and their simulation showed the possibility to produce such super Li-rich object based on the updated nuclear reaction rates and an asymmetric mixing model. It is noted that these suggestions should be related to the RGB bump because the extra mixing is inhibited by the mean molecular weight discontinuity left by the first dredge-up. After the outward-moving hydrogen burning shell erases the chemical discontinuity, the mixing is subsequently allowed to facilitate the CF mechanism, the resultant Li enhancement depends on the mixing speed and geometry \citep{Sackmann99}. Indeed, this Li enhancement site has been supported by the observations \citep{Charbonnel00}, this kind of mixing is likely to take effect on nine of our targets located around the RGB bump.

 A few Li-rich giants were reported beyond the RGB bump \citep{Monaco11, Martell13}, and two of our sample stars were also identified. In order to explain these Li-rich giants, \citet{Martell13} suggested that the thermohaline mixing would happen at any time past the RGB bump.  According to the calculation of \citet{Sackmann99}, Li would be decreased after the phase of Li enhancement because the freshly formed Li is destroyed again when the mixing goes backward to the high-temperature regions. Thus, the Li-rich giants beyond the RGB bump might be in the process of enriching Li or depleting Li again after the Li enhancement.

 Interestingly, several Li-rich giants were identified locating at the evolutionary stage of red clump based on the asteroseismic information \citep[e.g.][]{Kumar18, Carlberg15, Silva14}, which is drawing the attention of the community to understand the extra mixing happen at the RC phase. A large portion of our Li-rich giants belong to the red clump (see Table \ref{tab2}). Some work tried to reveal the physics related to Li-rich at this stage. According to the positions of a collected Li-rich sample in the HR diagram, \citet{Kumar11} found that some Li-rich giants might concentrate at the RC region, they suggested that the helium flash might trigger extra mixing to produce Li because all low-mass RC stars should experience the process of He flash. Based on the hydrodynamic simulation of the He flash, \citet{Mocak11} tried to explain the phenomenon of Li-rich RC star by the hydrogen injection, the injection of hydrogen into helium layer will induce an extra mixing, the matter including Li, C, and N that have been undergone the hydrogen fusion would be exposed to the surface by the mixing process. Nevertheless, \citet{Denissenkov12} noted that the episode of hydrogen injection flash would not work for the stars with a mass larger than 1.7 $M_{\odot}$. Alternatively, \citep{Casey19} speculated that the tidal interaction in a binary system would be responsible for Li production. The rotation-induced mixing can be enhanced with a factor of 6,500 by a binary companion, thus, a high ratio RC of Li-rich giants can be found.

 Thirteen of our sample stars are identified as the low-mass AGB stars. As the star evolved to AGB, the HBB principally operate to enrich Li for intermediate mass stars (4-8 $M_{\odot}$), but it is not expected for low-mass AGB stars \citep{Palmerini11}. This is because the bottom of the convective envelope of low mass AGB stars is too cool to trigger the CF mechanism. \citet{Ruchti11} discovered a few low-mass Li-rich AGB objects from the RAVE survey and suggested that extra mixing still take effect if the $^3$He is not fully depleted before AGB. It agrees with the simulation of \citet{Palmerini11}, they suggested that the Li abundance of low-mass AGB stars would be reasonable lower than that of RGB stars since the raw material of $^7$Li production, $^3$He, has been diluted before the AGB \citep{Nollett03, Uttenthaler07}. Indeed, the Li abundances show a systematic reduction as the star evolves past log $g \sim$ 2.0 (see Figure \ref{fig5}), which is the typical log $g$ of AGB star predicted by the evolutionary track for a 1.0 $M_{\odot}$ solar metallicity star. If we simply divided the Li-rich giants into before AGB and AGB by log $g \sim$ 2.0, the difference of mean A(Li) between these two groups is about 0.75 dex. Yet, we can not rule out the effect of selection bias on the A(Li) difference, because the selection of Li-rich candidate prefers the object with a large equivalent width of the Li line.

 Our results show that most of the newly found Li-rich giants cluster at the evolutionary stages of RGB bump, RC and AGB, and the RC and AGB stars occupy a large portion. It is suggested that there do not exist a universal scenario for stars at all evolution stages, multiple Li enriched mechanisms would be needed. Most of the Li-rich giants locate the evolutionary stages of RC and AGB at least for our sample, which may due to the long timescale of the two evolutionary phases.

\section{Summary} \label{sec5} We reported 44 newly Li-rich giants and homogeneously investigated them by IR excess, projected rotational velocity, stellar population, and evolutionary stage. With the advantage of high precision stellar parameters, the evolution stages of our Li-rich giants were well determined. Among them, we identified 20 RC stars, 13 AGB stars, nine RGB bump stars, and two stars just evolved past RGB bump. Moreover, eight of the twenty RC stars were reliably determined with both the HR diagram and the data-driven method of \citet{Ting18}. The non-unique evolutionary sites of Li-rich giants prefer that the various mechanisms of Li enhancement take effect on the different evolutionary stages.

 It is suggested that the Li-rich phenomenon associated with the IR excess and/or fast rotation are due to the external mechanism. In our sample, none of IR excess and only 6.8\% rate of fast rotator indicate that a minor contribution is caused by the external source. While the internal mechanism is mostly responsible for Li production considering both the Li abundance and the concentration in the HR diagram.

 It is hard to firmly distinguish the RC from the RGB bump without asteroseismic data, some discrepancies happen in our determination of evolutionary stages. Also, it is difficult to correct the selection bias, we can not tell the exact ratio among the different evolution stages for our targets. Therefore, it is essential to systematically investigate the Li-rich stars with the asteroseismic information from Kepler/TESS, etc.

\acknowledgments
 The authors thank the anonymous referee for useful comments to improve this paper. We also thank Jane Lin for helpful discussions on stellar mass for the Li-rich giants. This research was supported by National Key Basic Research Programme of China 2014CB845700, and by the National Natural Science Foundation of China under grant Nos. 11833006, 11603037 and 11473033. This work is supported by the Astronomical Big Data Joint Research Center, co-founded by the National Astronomical Observatories, Chinese Academy of Sciences and the Alibaba Cloud. K.P. acknowledges supports from the Mt. Cuba Astronomical Foundation Grant and from Center for Astronomical Megascience, CAS. Guoshoujing Telescope (the Large Sky Area Multi-Object Fiber Spectroscopic Telescope LAMOST) is a National Major Scientific Project built by the Chinese Academy of Sciences. Funding for the project has been provided by the National Development and Reform Commission. LAMOST is operated and managed by the National Astronomical Observatories, Chinese Academy of Sciences. This publication makes use of data products from the Wide field Infrared Survey Explorer, which is a joint project of the University of California, Los Angeles, and the Jet Propulsion Laboratory/California Institute of Technology, funded by the National Aeronautics and Space Administration. Based on data products from observations made with ESO Telescopes at the La Silla Paranal Observatory under programme ID 188.B-3002. This research made use of Astropy, a community-developed core Python package for Astronomy.

\software{Astropy \citep{Astropy13, Astropy18}, PARAM \citep{daSilva06}, SPECTRUM \citep{Gray94}, MARCS \citep{Gustafsson08}}

\appendix
\begin{longrotatetable}
\begin{deluxetable}{ccccccccccccccccll}
\tablecaption{Additional information of the Li-rich giants, including stellar parameters, Li abundances, mass, age, etc. \label{tab2}}
\tabletypesize{\scriptsize}
\tablewidth{0pt}
\tablehead{\colhead{ID} & \colhead{...}& \colhead{$T_\mathrm{eff}$}& \colhead{log $g$}& \colhead{[Fe/H]}& \colhead{vmic}& \colhead{A(Li)$_\mathrm{LTE}$}& \colhead{A(Li)$_\mathrm{NLTE}$}& \colhead{vsini}& \colhead{$T_\mathrm{eff}$$_\mathrm{ph}$}& \colhead{log $g_\mathrm{ph}$}& \colhead{...} & \colhead{Age}& \colhead{Mass}& \colhead{log(L/L$_\odot$)}& \colhead{$\chi_{[3.4],[22]}$}& \colhead{phase}}
\startdata
J005749.84+865043.5  & ... & 4596 $\pm$ 41 & 2.23 $\pm$ 0.13 &  0.26 $\pm$ 0.04 & 1.41 $\pm$ 0.06 & 2.06 $\pm$ 0.07 & 2.04 & 3.0 $\pm$ 0.4 &   4511 & 2.529 & ... & 2.136 & 1.761 & 1.722 $\pm$ 0.067 &  0.6 & II \\
J011727.43+461528.3  & ... & 4971 $\pm$ 55 & 2.67 $\pm$ 0.14 & -0.15 $\pm$ 0.06 & 0.92 $\pm$ 0.07 & 3.05 $\pm$ 0.09 & 2.97 & 3.7 $\pm$ 1.7 &   4599 & 2.425 & ... & 1.145 & 2.077 & 1.932 $\pm$ 0.075 & -0.8 & IV\tablenotemark{ab} \\
J012741.50+572857.3  & ... & 4569 $\pm$ 60 & 2.28 $\pm$ 0.24 & -0.33 $\pm$ 0.07 & 1.76 $\pm$ 0.10 & 3.32 $\pm$ 0.19 & 3.62 & 3.1 $\pm$ 0.5 &   4414 & 1.833 & ... & 7.071 & 1.073 & 2.170 $\pm$ 0.085 &  0.5 & V \\
J013000.68+561722.3  & ... & 4784 $\pm$ 54 & 3.33 $\pm$ 0.11 & -0.08 $\pm$ 0.05 & 1.11 $\pm$ 0.10 & 3.57 $\pm$ 0.13 & 3.40 & 3.2 $\pm$ 1.3 &   4701 & 2.607 & ... & 4.028 & 1.318 & 1.516 $\pm$ 0.070 &  0.3 & II \\
J013829.65+840530.5  & ... & 4784 $\pm$ 38 & 2.64 $\pm$ 0.10 &  0.11 $\pm$ 0.03 & 1.54 $\pm$ 0.06 & 3.60 $\pm$ 0.06 & 3.67 & 2.2 $\pm$ 0.3 &   4724 & 2.417 & ... & 1.786 & 1.835 & 1.887 $\pm$ 0.070 &  1.2 & IV\tablenotemark{a} \\
J023101.65+625705.3  & ... & 4587 $\pm$ 78 & 2.67 $\pm$ 0.24 &  0.04 $\pm$ 0.08 & 1.79 $\pm$ 0.11 & 2.79 $\pm$ 0.14 & 2.80 & 3.0 $\pm$ 0.5 &   4646 & 2.346 & ... & 7.770 & 1.123 & 1.716 $\pm$ 0.073 & -0.1 & II \\
J023607.44+445007.8  & ... & 4429 $\pm$ 74 & 2.17 $\pm$ 0.20 & -0.11 $\pm$ 0.08 & 1.62 $\pm$ 0.10 & 2.47 $\pm$ 0.14 & 2.52 & 3.5 $\pm$ 0.5 &   4268 & 1.972 & ... & 1.405 & 1.975 & 2.243 $\pm$ 0.076 &  0.3 & II \\
J024710.97+432606.0  & ... & 4315 $\pm$ 59 & 2.18 $\pm$ 0.18 & -0.16 $\pm$ 0.07 & 2.04 $\pm$ 0.11 & 3.24 $\pm$ 0.14 & 3.33 & 7.7 $\pm$ 0.6 &   4180 & 1.888 & ... & 6.847 & 1.127 & 2.056 $\pm$ 0.131 &  0.6 & III \\
\enddata
\tabletypesize{}
\tablenotetext{a}{The RC stars were classified by the position of HR-diagram without asteroseismic data.}
\tablenotetext{b}{The RC stars were identified by the data-driven method of \citet{Ting18}.}
\tablenotetext{ab}{The RCs were consistently classified by the two approaches.}
\tablecomments{The informations include the $T_\mathrm{eff}$ and log $g$ calculated by the photometric approach. The remaining columns are the IR excess index, stellar properties (i.e. the mass, age and evolutionary stage). The phase indicate the evolutionary stage of the Li-rich giants. The Roman numerals from I to V indicate the stages of before RGB bump, the RGB bump, above RGB bump, RC and AGB. A portion of Table 2 is shown here for guidance regarding its form and content, the full table is available at the CDS.}
\end{deluxetable}
\end{longrotatetable}

\end{document}